\documentclass[aps,prl,reprint,groupedaddress,showpacs]{revtex4-1}
\usepackage{amsmath}
\usepackage{graphicx}
\usepackage{longtable}
\RequirePackage{ifthen}
\usepackage[usenames,dvipsnames]{color}

\begin{document}
 \marginparwidth 2.7in
\title{Coupled nonpolar-polar metal-insulator transition in 1:1 SrCrO$_3$/SrTiO$_3$ superlattices: A first-principles study}


\author{Yuanjun Zhou}
\affiliation{Rutgers, the State University of New Jersey}
\author{Karin M. Rabe}
\affiliation{Rutgers, the State University of New Jersey}


\date{\today}

\begin{abstract}
Using first principles calculations, we determined the epitaxial-strain dependence of the ground state of the 1:1 SrCrO$_3$/SrTiO$_3$ superlattice. The superlattice layering leads to significant changes in the electronic states near the Fermi level, derived from Cr $t_{2g}$ orbitals. 
An insulating phase is found when the tensile strain is greater than 2.2\% relative to unstrained cubic SrTiO$_3$. The insulating character is shown to arise from Cr $t_{2g}$ orbital ordering, 
which is produced by an in-plane polar distortion that couples to the superlattice d-bands and is stabilized by epitaxial strain.
This effect can be used to engineer the band structure near the Fermi level in transition metal oxide superlattices.
\end{abstract}



\maketitle

\def\scr{\scriptsize}
\def\draftversion{true}
\ifthenelse{\equal{\draftversion}{true}}{
  \marginparwidth 2.7in
  \marginparsep 0.5in
  \newcounter{comm} 
  \def\commnext{\stepcounter{comm}}
  \def\commtext{{\bf\color{blue}[\arabic{comm}]}}
  \def\commmar{{\bf\color{blue}[\arabic{comm}]}}
  \def\yj#1{\commnext\marginpar{\small YJZ\commmar: #1}\commtext}
  \def\krm#1{\commnext\marginpar{\small KMR\commmar: #1}\commtext}
  \def\tnewpage{\marginpar{\small Temporary newpage}\newpage}
}{
  \def\yj#1{}
  \def\krm#1{}
  \def\tnewpage{}
}

Metal-insulator transitions (MIT) occur in many transition metal oxides as a function of composition, temperature, pressure and epitaxial strain \cite{MITrev}.
At the atomic scale, the mechanism for the metal-insulator transition depends on the physics of the insulating state. For Mott insulators, the relevant parameters are the transition metal (TM) $d$-level occupation, which can be changed through doping, and the bandwidth, which can be changed by varying the TM-oxygen-TM bond angles. For band insulators, a gap can be opened at the Fermi level by a structural distortion that lowers the crystal symmetry. A symmetry-breaking distortion that lifts the degeneracy of a partially occupied state will always lower the energy; this is called a Jahn-Teller distortion or a Peierls distortion depending on whether the degenerate states are localized atomic orbitals or extended bands. Polar distortions have also been shown to couple to states near the Fermi level, proving useful in band-gap engineering\cite{Neaton-STO-bandeng,Rappe-bandeng}. 
When the distortion can be controlled by an applied field or stress, the system can dynamically be driven back and forth through the MIT, a property of particular interest for applications to high-performance switching devices\cite{MITappkim,MITtransistor},

SrCrO$_3$ (SCO) is a $d^2$ perovskite. 
It was initially reported as a paramagnetic and metallic cubic perovskite\cite{SCOfirst}.
Experimental studies also showed a tetragonal C type antiferromagnetic (C-AFM) phase, with the Neel temperature below 100K and space group $P4/mmm$, coexisted with the cubic phase in the low temperature\cite{ACrO3,OSMartin}. 
This tetragonal C-AFM state is further investigated in first-principles studies\cite{Lee-Pickett,SCO-Zhu}.
It is correlated to a partial orbital ordering $d_{xy}^1$,$(d_{yz}d_{xz})^1$, where the $d_{xy}$ is nearly occupied while $d_{yz}$ and $d_{xz}$ orbitals are both half occupied\cite{OSMartin,Lee-Pickett}. 
For the single layer film of SCO epitaxially grown on SrTiO$_3$, 
due to the missing apical oxygen of the CrO$_5$ octahedra at the surface, $d_{yz}$ and $d_{xz}$ orbitals become lower in energy than $d_{xy}$, leading to a $d_{yz}^1d_{xz}^1d_{xy}^0$  orbital ordering and an insulating state\cite{SCOferro}.
However, recently Zhou {\it et al} claimed that SCO is an insulator while the insulator-metal transition occurs under sufficient pressure\cite{Zhou-Goodenough}, due to the bond instability found around 4 GPa, suggesting the strong interaction between structure and electronic structure. 

Layering a $d$-band perovskite oxide in a superlattice with a $d^0$ insulating perovskite will lead to large changes in the $d$-band states near the Fermi level, depending on the thickness of the $d$-band layer \cite{SFOSTO-James,SROSTO-Javier,ZWY-SrCoO3SrTiO3}. 
For SCO, SrTiO$_3$ (STO) is a natural choice as the second component. It is a $d^0$ insulator with a 3.25 eV band gap and an A-site cation in common with SCO. In addition, the structure of STO can be tuned by epitaxial strain, with different octahedral rotation patterns and polar distortions in compressive and tensile strain\cite{STO-Antons,STO-oswaldo,stacking}; these distortions are expected to influence the structure in the SCO layer.

In this Letter, we present first-principles results for the ground-state phase sequence of the 1:1 SrCrO$_3$/SrTiO$_3$ superlattice with varying epitaxial strain, with a first-order MIT from a metallic x-type $P2_1/c$ phase to a polar insulating G-type $Pmm2$ phase observed at 2.2\% tensile epitaxial strain. 
We show that the insulating character arises from orbital ordering induced by 
a polar distortion in the SCO layer.
The polar distortion is shown to be stabilized by a combination of epitaxial strain and the polar distortion in the adjacent STO layer.
This offers the possibility of driving the system through the MIT by an applied field or stress that coupled to the polar mode.

Our first-principles calculations were performed using the local density approximation\cite{LDA1,LDA2} with Hubbard U (LDA+U) method implemented in the $Vienna$ $Ab$ $initio$ $Simulation$ $Package$ (VASP-5.2\cite{vasp1,vasp2}). 
We used the Dudarev implementation\cite{Dudarev} with effective on-site Coulomb interaction $U=1.5$ eV to describe the localized 3$d$ electron states of Cr atoms. The projector augmented wave (PAW) potentials\cite{paw,paw2} contain 10 valence electrons for Sr ($4s^24p^65s^2$), 12 for Cr ($3p^63d^44s^2$)10 for Ti ($3p^63d^24s^2$) and 6 for O ($2s^22p^4$). $\sqrt 2\times\sqrt 2 \times 2$ and $2\times 2\times 2$ cells were used to allow the structural distortions and magnetic orderings for the 1:1 superlattice. 500 eV energy cutoff, $4\times 4\times 4$ k mesh and $5\times 10^{-3}$  eV/\AA~force threshold  were used for structural relaxations. 
A $15\times 15\times 11$ k-point mesh was used for density of state calculations.  
Our calculations give $a=3.812$\AA~for the relaxed tetragonal state of C-AFM SrCrO$_3$, consistent with the previous first-principles and experimental studies\cite{Lee-Pickett,OSMartin}, with aspect ratio $c/a =0.964$ slightly smaller than the experimental value of 0.99. 
The epitaxial strain value is defined with respect to 3.849\AA, the computed lattice constant of the cubic STO.
The effects of epitaxial strain were studied through ``strained bulk" calculations\cite{strainedbulk}. 

At each value of epitaxial strain in the range $-4\% $ - $3\%$, we determined the ground state (GS) structure of the superlattice using the ``stacking method'' \cite{stacking}. This involved combining the computed low-energy structures of epitaxially strained STO and SCO to obtain a set of starting structures, which were then relaxed to the nearest energy minimum using first-principles calculations. 
To obtain the low-energy structures of SCO, we considered F, G, A, and C-type AFM ordering in the tetragonal phase with various distortions.
C-AFM was found to be the ground state magnetic ordering in the full range of epitaxial strain considered. For C-AFM and FM states, $P4/mmm$ is the GS structure. 
For G-AFM, $P4/mmm$ is the GS from $-4\%$ to 0\%,  $P4/mbm$ Jahn-Teller (JT)\cite{JT, Junhee-JT} distortion is the GS beyond $0\%$ and anti-polar distortion $Pmcm$ is also unstable beyond 2\%. 
For A-AFM, $P4/mmm$ is the GS from $-4\%$ to 0\%, $P4/mbm$ Jahn-Teller (JT)\cite{JT, Junhee-JT} distorted structure is the GS beyond $0\%$, and polar distortion $Pmm2$ is the unstable beyond 3\%.
The low-energy structures of epitaxially-strained STO have been previously discussed in the literature\cite{STO-Antons,STO-oswaldo,stacking}.
We use the low-energy structures of STO listed in Table I of Ref.~\cite{stacking}, consistent with the reports in other studies.
When the epitaxial strain is within $-1\%$ - $1\%$, the low-energy structures of STO are generated by the M and R point octahedral rotations.
For compressive epitaxial strain greater than $-1\%$, the out-of-plane polar mode becomes unstable, while the in-plane rotations are suppressed. 
For tensile epitaxial strain greater than 1\%, the main distortions are in-plane polar modes along [110] and [100] and out-of-plane rotations are suppressed. 

 \begin{figure}
 \includegraphics[width=0.5\textwidth]{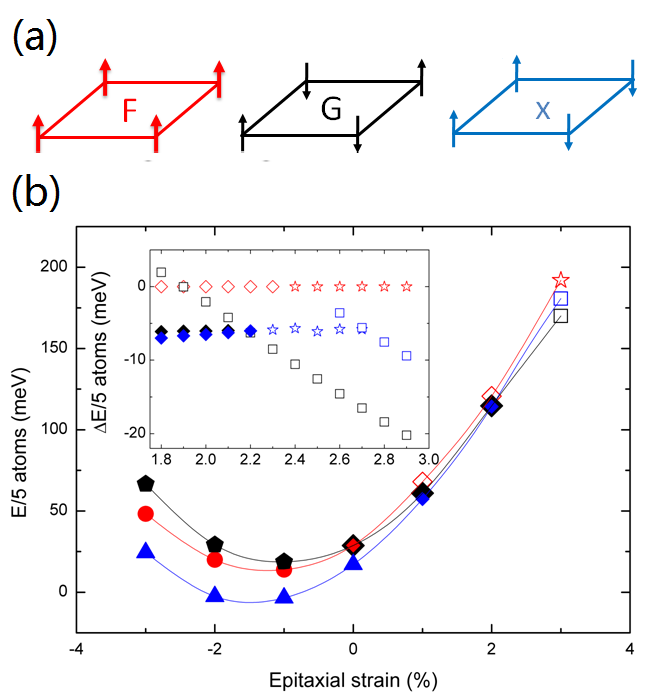}%
 \caption{\label{structure}(a) Magnetic orderings in the SCO layer. Red: ferromagnetic (F), black: G-AFM (G), and blue: x-type AFM (x) states.
(b) 
GS structures and energies for F (red), G (black), and x (blue) magnetic states as functions of epitaxial strain.
The solid curves guide the eye. 
Insulating and metallic states are denoted by open and solid symbols.
Shapes of data points indicate the space groups: pentagons ($Pbam$), triangles ($P2/m$), circles ($P4/mbm$), diamonds ($P2_1/c$), five-pointed stars ($Pc$) and squares ($Pmm2$). 
The inset shows the energies of low energy structures relevant to the F state in the epitaxial strain range 1.8\% to 2.9\%.}  
 \end{figure} 

To include magnetic ordering in the stacking method
for the 1:1 superlattice, we consider three magnetic phases for the single layer of SCO that can fit in the $2\times 2$ lattice, as shown in Fig.~\ref{structure}(a). The three phases are FM, G-AFM for the checkerboard ordering in the CrO plane, and x-type AFM for the AF ordering along [100] and FM along [010] in the CrO plane.
For each magnetic state of SCO and each value of epitaxial strain, we generate the starting structures by following the procedures of the stacking method. 
We also carry out relaxations for a small number of random starting structures\cite{stacking}. 
Thes show that from $-3\%$ to $-1\%$ epitaxial strain, the JT distortion is found in the SCO layer, while it is not metastable in the bulk state.

In Fig.\ref{structure}(b) we plot the total energies of the GS of the superlattice obtained by the ``stacking method''  for the three magnetic states versus epitaxial strain. 
From $-3\%$ to $2.2\%$, the x-AFM is the GS magnetic state, as expected from the resemblance of magnetic ordering to the bulk C-type AFM state.
Beyond 2.2\%, the GS magnetic state changes to G-AFM.

It is instructive to examine the epitaxial strain dependence of the structure for each magnetic ordering. 
For FM state, 
from $-3\%$ to $0\%$, $P4/mbm$ structure with the $a^0a^0c^-$ oxygen octahedral rotation pattern is the lowest energy structure in this phase.
From $0\%$ to 2.3\%, the lowest energy structure is $P2_1/c$, with an $a^-a^-c^-$ rotation pattern and JT $Q_2$ mode. The amplitude of the $Q_2$ mode is about 1\% of the lattice constant and does not change much with the strain. As the tensile epitaxial strain increases, the $a^0a^0c^-$ rotation decreases while the $a^-a^-c^0$ increases.
Beyond 1\%, the lowest energy structure of FM state is insulating. 
Beyond 2.3\%, the lowest energy structure is $Pc$, with the $a^-a^-c^0$ rotation and small and decreasing amount of $a^0a^0c^-$ rotation, JT $Q_2$ mode in the SCO layer, and polar distortions along [110].

For the G-AFM state,
from $-3\%$ to $-1\%$, $Pbam$ is the lowest energy structure, with the $a^0a^0c^-$ rotation mainly in the STO  and JT $Q_2$ distortion in the SCO layer. The JT distortion amplitude is small but enough to lift the $d_{yz}$ $d_{xz}$ degeneracy. However, the system is still metallic due to the band overlap.
From 0\% to 2.2\%, $P2_1/c$ with the  $a^-a^-c^-$ oxygen octahedral rotation pattern and JT $Q_2$ mode is the lowest energy structure.
Beyond 2.2\%, the polar structure $Pmm2$, with only polar distortion along [100] becomes the lowest energy structure, leading to the insulating overall GS, with magnetic moment 1.8 $\mu_B$.

For the x-AFM state,
from $-3\%$ to $0\%$, the lowest energy structure is nonpolar $P2/m$, with $a^0a^0c^-$ rotation mainly in the STO layer, and small JT $Q_2$, $Q_3$ modes in the SCO layer.
From 0\% to 2.2\%, the lowest energy structure is $P2_1/c$.
Beyond 2.2\%, the lowest energy structure is insulating. 
In particular, from 2.2\% to 2.7\%, the lowest energy structure is $Pc$ with the  $a^-a^-c^0$ rotation and a small and decreasing amount of $a^0a^0c^-$ rotation, JT $Q_2$ mode in the SCO layer and polar distortion along [110]. The in-plane rotation is now the main distortion pattern, and the amplitudes for both STO and SCO layers are similar, due to the connection of oxygen octahedra.
Beyond 2.8\%, the lowest energy structure changes to $Pmm2$.

The phase boundary at 2.2\% epitaxial strain is of particular interest.
The inset of Fig.~\ref{structure} shows the energies for F, G, and x structures for $1.8\% - 2.9\%$ epitaxial strain. The energy of the G-type polar $Pmm2$ structure decreases relative to that of the x-type nonpolar $P2_1/c$ structure with increasing epitaxial strain, with a first-order transition at 2.2\% epitaxial strain. We note the metal-insulator transition associated with the change from the nonpolar to the polar structure.

\begin{figure}
 \includegraphics[width=0.5\textwidth]{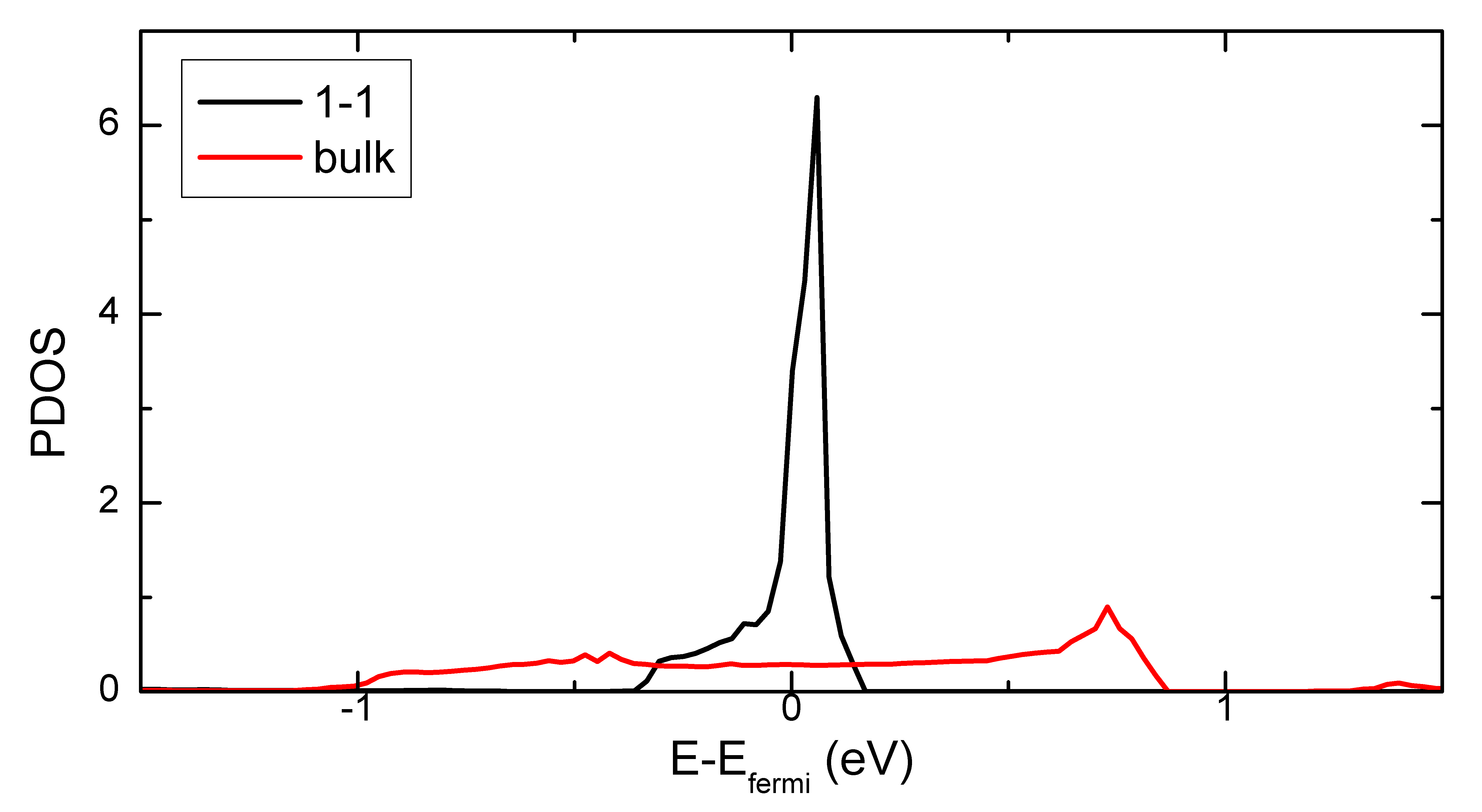}%
 \caption{\label{hybrid}
 The PDOS of the Cr $d_{yz}$ orbital for G-type 1:1 SCO/STO superlattice and C-type bulk SCO. }
 \end{figure}

To figure out the mechanism of the insulating phase in the 1:1 superlattice beyond 2.2\% epitaxial strain, we first consider the effect of the superlattice layering on the bands near the Fermi level.
It eliminates the dispersions of Cr $t_{2g}$ bands along $k_z$, and significantly narrows the band widths of $d_{yz}$ and $d_{xz}$.
Fig.~\ref{hybrid} shows the PDOS of $d_{yz}$ of one Cr atom in the G-type 1:1 superlattice and C-AFM bulk SCO for 3\% tensile epitaxial strain in the $P4/mmm$ symmetry. 
The 1:1 superlattice is metallic for the $P4/mmm$ structure, with the $d_{yz}$ band much narrower than  in the bulk case.
The thickness of the STO layer has little effect. Our calculations show that the Cr-layer structure and bands are very similar for 1:1 and 1:3 superlattices.

\begin{figure}
 \includegraphics[width=0.5\textwidth]{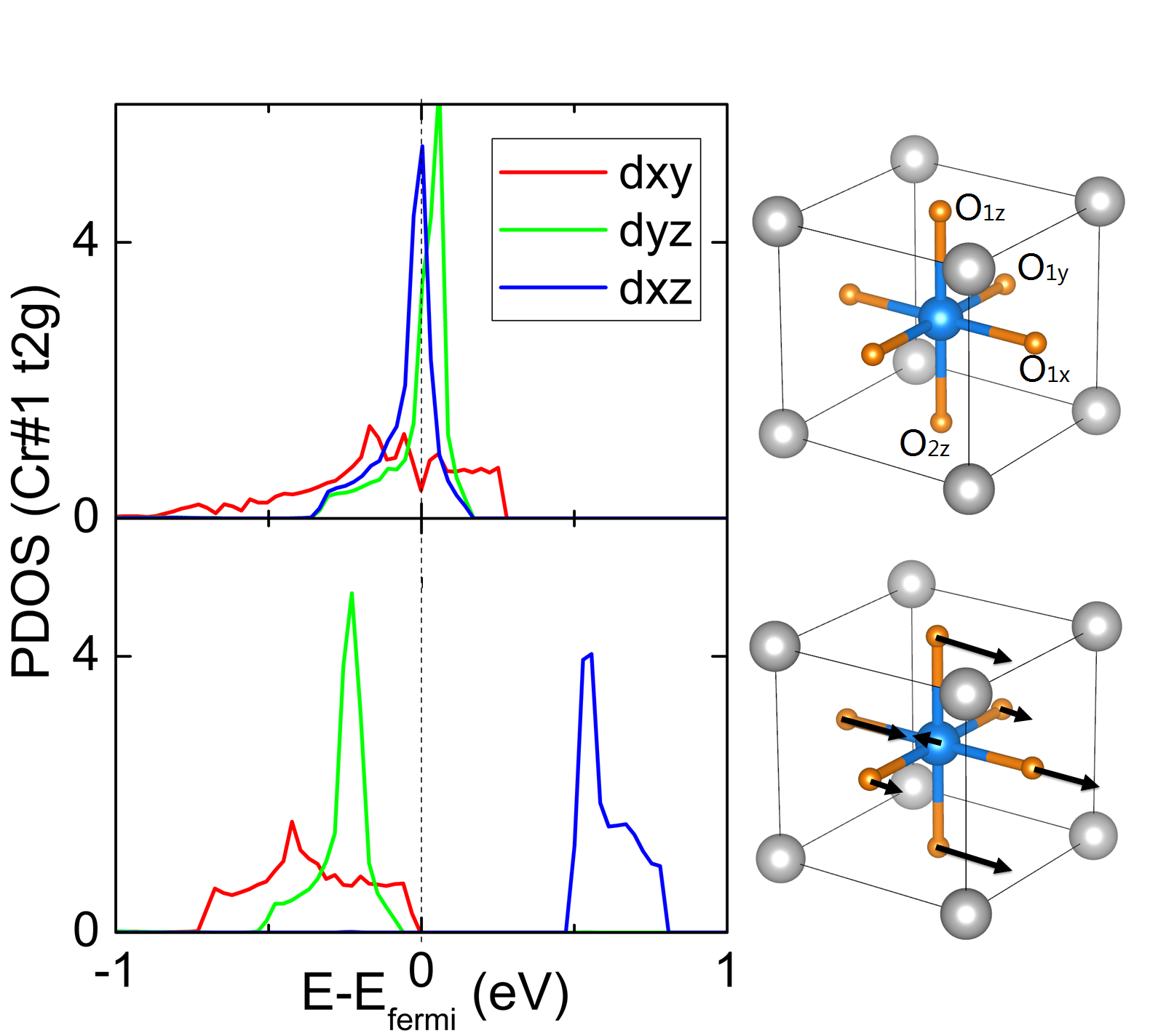}%
 \caption{\label{u00a00+3}PDOS of the spin up Cr $t_{2g}$ in the 1:1 superlattice for +3\% strain, with G-AFM magnetic ordering. The left panels represent the PDOS without (top) and with (bottom) the polar distortion. The vertical dashed line marks the energy of the highest occupied state. The distortions in the SCO layer are shown on the right part, where gray, blue and orange spheres represent Sr, Cr and O ions, respectively.}
 \end{figure}
 

Next, we consider the coupling effect of the polar distortion to the superlattice insulating $Pmm2$ state.
The $Pmm2$ structure is generated from the high symmetry reference structure by the doubly-degenerate in-plane $E_u$ polar distortion, with eigenvector
 (0.00,  0.00,  $-$0.15, 0.08, 0.53,
       0.53, 0.37, 0.40, 0.06, 0.33)
specifying the displacement pattern of the atoms (Sr, Sr, Cr, Ti, O$_{z1}$, O$_{z2}$,
O$_{x1}$, O$_{x2}$, O$_{y1}$, O$_{y2}$). 
From 2.2\% to 3\%, the band gap increases from 0.27 to 0.47 eV, 
and the polarization increases from 36 to 41 $\mu C/cm^{-2}$.

The effects of the in-plane polar distortion on the electronic structure of the superlattice are evident from the projected density of states (PDOS) of the $t_{2g}$ bands of the spin up Cr atom, shown in Fig.~\ref{u00a00+3}.
The layered structure of the superlattice splits $d_{xy}$ from $d_{xz}$ and $d_{yz}$ and increasing tensile epitaxial strain lowers the energy of the $d_{xy}$ orbital relative to $d_{yz}$ and $d_{xz}$.
At 3\% epitaxial strain, in the undistorted structure, all three $d$ orbitals are partially occupied. The in-plane polar distortion lifts the degeneracy of $d_{yz}$ and $d_{xz}$, so that in the polar $Pmm2$ state the $d_{xy}$ and $d_{yz}$ orbitals are fully occupied while $d_{xz}$ is unoccupied, corresponding to $d_{xy}^1d_{yz}^1d_{xz}^0$ orbital ordering. 

\begin{figure}
 \includegraphics[width=0.5\textwidth]{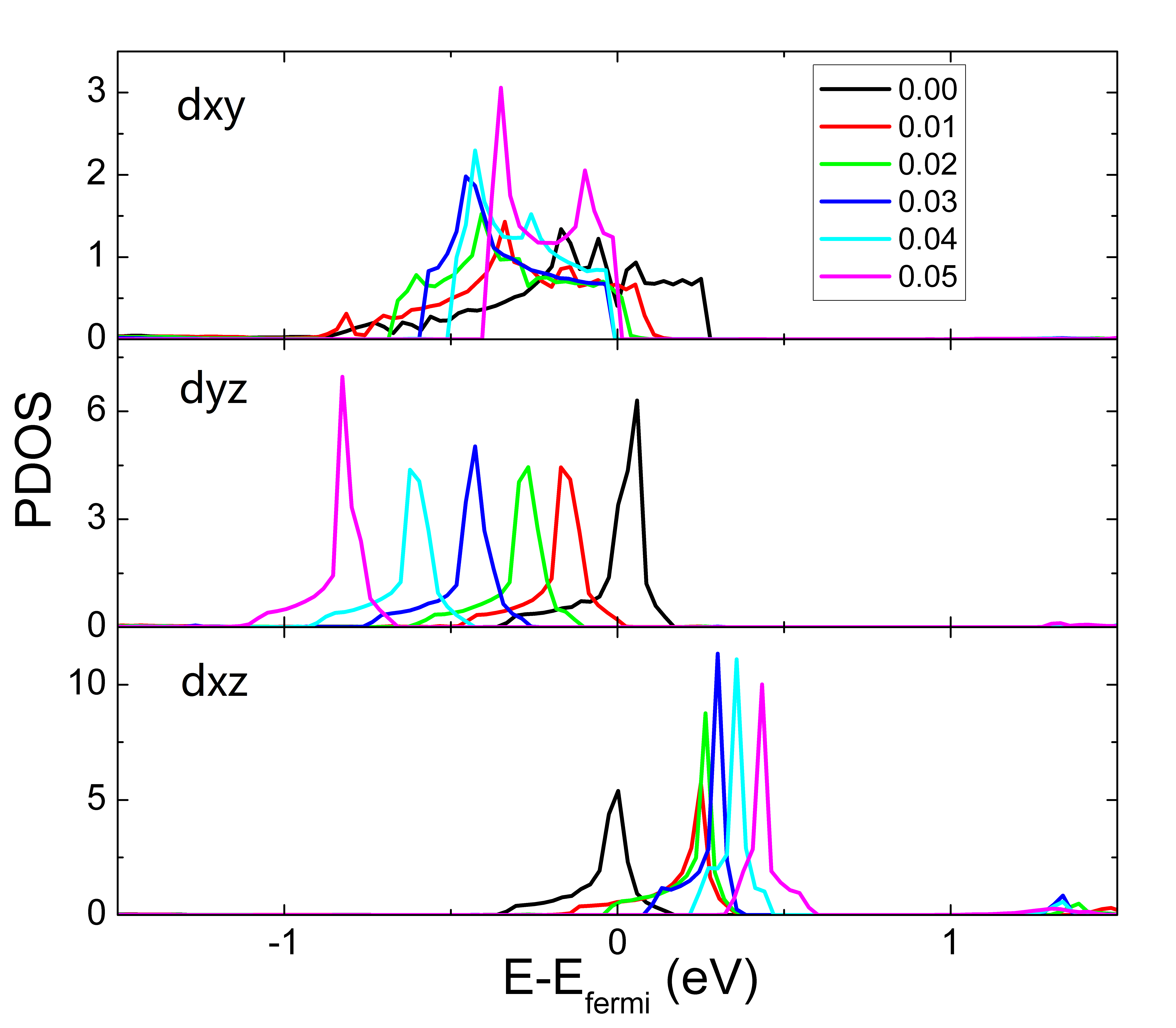}%
 \caption{\label{Pmm2path}The evolution of PDOS of Cr $d_{xy}$, $d_{yz}$ and $d_{xz}$ as a function of polar displacement of Cr ion in [100]. The amplitude of the displacement for each curve is given in the legend in units of the in-plane lattice constant.}
 \end{figure}

In fact, the OO can be produced by polar displacements just of the Cr atoms.
To show this, we consider a $Pmm2$ 1:1 SCO/STO superlattice, in which only the Cr atoms are uniformly displaced along [100], while all other atoms stay at the high symmetry positions. In Fig.~\ref{Pmm2path} we show the PDOS of the distorted superlattice as a function of Cr displacement.
There is a dramatic downward shift of the $d_{yz}$ with increasing displacement, accompanied by a smaller upward shift of $d_{xy}$ and a decrease in the $d_{xy}$ bandwidth.

Finally, we discuss why the in-plane polar distortion becomes the GS for tensile strain.
Our calculations show that for any distortion that lifts the $d_{yz}$ and $d_{xz}$ degeneracy, there is a linear energy gain. This explains the emergence of JT distortions in $P2_1/c$ or $Pbam$ structures for the 1:1 superlattice.
However, in general, for large tensile epitaxial strain, the in-plane polar state is likely to be more favorable than the JT distortion due to the well-known polarization-strain coupling. 
The polar distortion in the STO layer is also an important influence.
To show this, we fixed the STO layer in a nonpolar structure for 3\% strain, froze in the in-plane polar mode and JT distortion in the CrO$_2$ layer in turn, and relaxed. Distortions survive in both cases and the relaxed structure with JT distortion has lower energy, suggesting that the in-plane polar mode in the STO coupled differently to the in-plane polar mode and the JT mode in the SCO layer.
   
This insulating state also raises the possibility of controlling band gap by applied electric field.
Given the SCO/STO superlattice in the insulating state, an in-plane electric field will change the atomic positions, and hence change the band gap, because the band gap is sensitive to the displacement of the Cr atom relative to the O atoms around it.  

In summary, we have studied both the lattice and electronic structures of the ground state for the 1:1 SCO/STO superlattice. Distortions in SCO layers are established by the superlattice layering with STO. 
For tensile epitaxial strain, due to the in-plane polar distortion associated with nonzero Cr displacements, the $d_{xy}^1d_{yz}^1d_{xz}^0$ orbital ordering can be formed and the band gap is therefore induced.
The polar distortion induced metal-insulator transition can be used to engineer the SCO band structures near the Fermi level. 
Our study sheds light on a new way to control electronic band structures and approach the metal-insulator transition point. 

We acknowledge helpful discussions with K. Garrity, D. R. Hamann, H. Huang, S. Y. Park, D. Vanderbilt and H. Zhang. First-principles calculations are performed on the Rutgers University Parallel Computer (RUPC) cluster and the Center for Functional Nanomaterials (CFN) cluster at Brookhaven National Lab. This work is supported by NSF DMR-1334428, ONR N00014-11-1-0665, and ONR N00014-11-1-0666.

\bibliography {SCOSTO}
\end{document}